\documentclass[usegraphicx,usenatbib,useAMS]{mn2e}
\newcommand\lsim{\mathrel{\rlap{\lower4pt\hbox{\hskip1pt$\sim$}}
        \raise1pt\hbox{$<$}}}
\newcommand\gsim{\mathrel{\rlap{\lower4pt\hbox{\hskip1pt$\sim$}}
        \raise1pt\hbox{$>$}}}

\usepackage{amsmath}
\usepackage{array}

\usepackage{graphicx}
\usepackage{fixltx2e}

\title[${H_2}~rovibrational~populations~and~self-shielding$]{${\bf H_2}$ self--shielding with non-LTE rovibrational 
populations: implications for cooling in protogalaxies} 
\author[J. Wolcott-Green et al.]
{J. Wolcott-Green and Z. Haiman \thanks{E-mail: jemma@astro.columbia.edu;
zoltan@astro.columbia.edu}\\
Department of Astronomy, Columbia University, 550 West 120th Street, MC 5246, New York, NY 10027, USA}
\begin{document}

\date{}

\pagerange{\pageref{firstpage}--\pageref{lastpage}} \pubyear{2012}

\maketitle

\label{firstpage}

\begin{abstract}

The abundance of molecular hydrogen (${\rm H_2}$), the primary 
coolant in primordial gas, is critical for the thermodynamic 
evolution and star--formation histories in early protogalaxies. 
Determining the photodissociation rate of ${\rm H_2}$ by an 
incident Lyman-Werner (LW) flux is thus crucial, but prohibitively 
expensive to calculate on the fly in simulations. The rate is 
sensitive to the ${\rm H_2}$ rovibrational distribution, which 
in turn depends on the gas density, temperature, and incident LW 
radiation field. We use the publicly available {\sc cloudy} package 
to model primordial gas clouds and compare exact photodissociation 
rate calculations to commonly--used fitting formulae. We find the 
fit from \citet{WGHB11} is most accurate for moderate densities 
${\rm n \sim 10^3 cm^{-3}}$ and temperatures, $T \sim 10^3$K, 
and we provide a new fit, which captures the increase in the rate 
at higher densities and temperatures, owing to the increased excited 
rovibrational populations in this regime. Our new fit has typical 
errors of a few percent percent up to ${\rm n\leq 10^7~cm^{-3}}$, 
${\rm T \leq 8000}$K, and ${\rm H_2}$ column density ${\rm N_{H2} 
\leq 10^{17}~cm^{-2}}$, and can be easily utilized in simulations. 
We also show that pumping of the excited rovibrational states of 
${\rm H_2}$ by a strong LW flux further modifies the level
populations when the gas density is low, and noticeably decreases
self-shielding for $J_{21} > 10^3$ and ${\rm n < 10^2 cm^{-3}}$. 
This may lower the ``critical flux'' at which primordial gas remains 
${\rm H_2}$--poor in some protogalaxies, enabling massive black hole 
seed formation.
\end{abstract}


\section{Introduction}
It has long been known that the cooling of metal--free primordial
gas, from which the first generation of stars form in protogalaxies, 
is dominated by ${\rm H_2}$ molecules \citep{Saslaw+Zipoy}. Once these 
first (``Population III'') stars begin to shine, however, the UV 
radiation they emit begins to photodissociate ${\rm H_2}$ via the 
Lyman--Werner (LW) bands (${\rm E_\nu =11.1-13.6 eV})$. 
Photodissociation feedback is non--trivial to model, particularly 
when the ${\rm H_2}$--column density becomes sufficiently high for 
it to become self--shielding (${\rm N_{H_2}}\gsim 10^{13} 
{\rm cm}^{-2})$. Accurate modeling is important because the 
thermodynamic and star formation histories in early protogalaxies 
depend sensitively on the ${\rm H_2}$ abundance. 

An example of this sensitivity occurs in the well--known 
``direct collapse'' scenario, in which a protogalaxy may 
avoid fragmentation (and thereby Pop III star formation) 
if exposed to a sufficiently strong UV flux from a near 
neighbor \citep[e.g.][and references therein]{Regan+17}. 
In this case, the ${\rm H_2}$--abundance remains too low 
to cool below the virial temperature of the halo, preventing 
fragmentation on stellar scales. It has been proposed that 
the resulting rapid accretion possible in this scenario may 
lead to the formation of massive black holes (${\rm M_{BH} 
\gsim 10^{4-5} M_\odot}$) that seed the earliest quasars 
\citep[see reviews by][]{VolonteriRev10,SMBHreview13,WiseRev18}. 
The predicted ``critical'' UV--flux to keep protogalactic 
[gas ${\rm H_2}$--poor (commonly denoted $J_{\rm crit}$) 
depends sensitively on the detailed calculation of the 
optically--thick ${\rm H_2}$--photodissociation rate. 

Calculating the full optically--thick photodissociation rate 
on the fly is prohibitively expensive in simulations due 
to the large number of LW transitions. There are a total 
of 301 rovibrational states of the ground electronic state (X) 
and over half a million allowed electronic transitions. 
Previous studies have relied, therefore, on fitting formulae 
for the optically--thick ${\rm H_2}$ photodissociation rate 
provided by \citet{DB96}. Their fit models the behavior well 
when primarily the ground (ortho and para) states of the 
molecule are populated, Alternatively, some studies use the 
modifed fit provided by \citet[][hereafter WGHB11]{WGHB11}, 
which is more accurate at higher densities and temperatures, 
when the rotational levels of the ground vibrational state 
are in, or close to, LTE.  

Both of these approximations are gross simplifications of the 
true rovibrational level populations, which in general are 
time-dependent and sensitive to the gas density, temperature, 
and rate of UV excitation. As discussed in WGHB11, the 
optically--thick rate can be quite sensitive to changes in 
the level populations, and in particular to the number of 
states that contribute to self--shielding, as more states 
becoming signficantly populated reduces the effective 
self--shielding column density. 

In this paper, we use the publicly available 
{\sc cloudy}\footnote{www.nublado.org} package \citep{Ferland+17} 
to calculate the ${\rm H_2}$ rovibrational populations under 
conditions similar to those in a pristine protogalactic gas 
cloud irradiated by UV. We compare the resulting optically--thick 
photodissociation rate to that predicted by the commonly used 
fitting formulae. We find that the fitting formula provided 
by WGHB11 is accurate in a narrow range of densities and temperatures 
(${\rm n \approx 10^3 cm^{-3}}$ and a few$\times 10^3$K), and 
{\it we provide an improved fitting formula} that is accurate 
for a larger swath of the parameter space. In particular, our 
improved fit matches the true rate of only a few percent up to 
$T=8000$K, ${\rm n=10^7~cm^{-3}}$, ${\rm N_{H2}=10^{16}~cm^{-2}}$, 
and with typical errors of order ten per cent up to higher 
column density, ${\rm N_{H2} = 10^{17}~cm^{-2}}$.

In the case of a protogalactic candidate for direct collapse, 
there is an additional modfication to the rovibrational 
distribution and thus to the optically--thick ${\rm H_2}$--
photodissociation rate that may be important, and which has 
not been considered in this context previously. When 
irradiated in the UV, decays following electronic excitation 
populate excited rovibrational levels of the ground electronic 
state (X), which subsequently decay through infrared emission 
in a radiative cascade. In the presence of a very strong UV 
flux, the radiative cascade of the UV--pumped ${\rm H_2}$ 
molecules can be interrupted by absorption of another UV photon. 
\citet{Shull78} found this ``re-pumping'' affects the radiative 
cascade $J_{21} \gsim 10^3$, and is {\it more likely} than 
decay of the molecule to a lower rovibrational state when the 
incident UV flux exceeds $J_{21} \sim 10^5$, with $J_{21}$ 
defined in the usual way: $J_\nu = J_{21} \times 10^{-21}
{\rm ~erg~s^{-1}~cm^{-2}~sr^{-1}~Hz^{-1}}$.

We find in our \textsc{cloudy} models that pumping from 
excited rovibrational states, leading to reduced effective 
column density, can decrease self--shielding by up to an 
order of magnitude at fluxes as low as $J_{21}=$ a few 
$\times 10^3$ when the density is low ${\rm n<10^3~cm^{-3}}$. 
Importantly, this is similar to the common determinations 
of $J_{\rm crit,21}\sim 10^{3}$ (e.g. WGHB11). At higher 
densities ($>10^3~{\rm cm^{-3}}$), the rovibrational 
level populations tend toward their LTE values and we find 
there is no effect on the self--shielding behavior even 
for the strongest UV flux we consider, $J_{21} \sim 10^5$. 

This paper is organised as follows: In \S~\ref{Sec:Model} 
we describe the details of our numerical modeling; in 
\S~\ref{Sec:Results} we discuss our results using a variety 
of \textsc{cloudy} models, and an updated fitting formulae 
for the optically--thick ${\rm H_2}$ photodissociation rate.  
We summarize our results and offer conclusions in 
\S~\ref{Sec:Conclusions}.

\section{Numerical Modeling}
\label{Sec:Model} 
\begin{figure*}
  \includegraphics[clip=true,trim=0.0in 3.3in 0in 0in,
    height=3.in,width=5.3in]{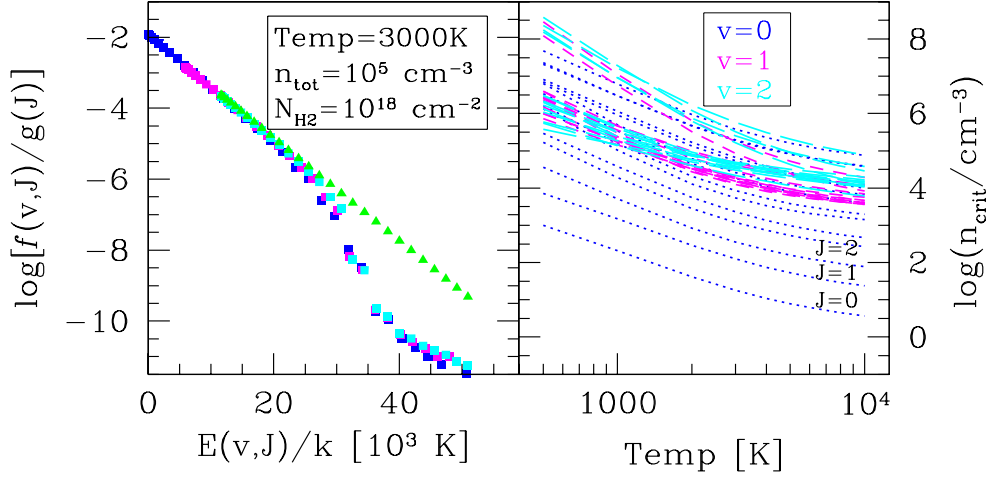}
  \caption{{\bf Left} The fractional populations of ${\rm H_2}$ in
rovibrational states with ${\rm v=0-2}$ and energy ${\rm E_{v,J}}$
predicted by the {\sc cloudy} ``Large ${\rm H_2}$'' model. 
Green triangles show the LTE populations while the resolved 
(non--LTE) populations in ${\rm v=0,1,2}$ are shown by dark blue, 
magenta, and cyan squares, respectively. 
{\bf Right}: The temperature dependence of the critical density 
(for LTE) is shown for each of the states ${\rm v=(0-2),J}$ 
(with dotted, short--dash, and long--dash lines, respectively.)
As illustrated in this panel, $T=3,000$K, ${\rm n=10^5~cm^{-3}}$ 
exceeds the critical density of the lowest $\sim 35$ states, and
the populations of these states, shown in the left--hand panel,
are indeed near their LTE values, as expected.}
  \label{Fig:ncrit}
\end{figure*}
\begin{figure*}
  \includegraphics[height=5.8in,width=7in]{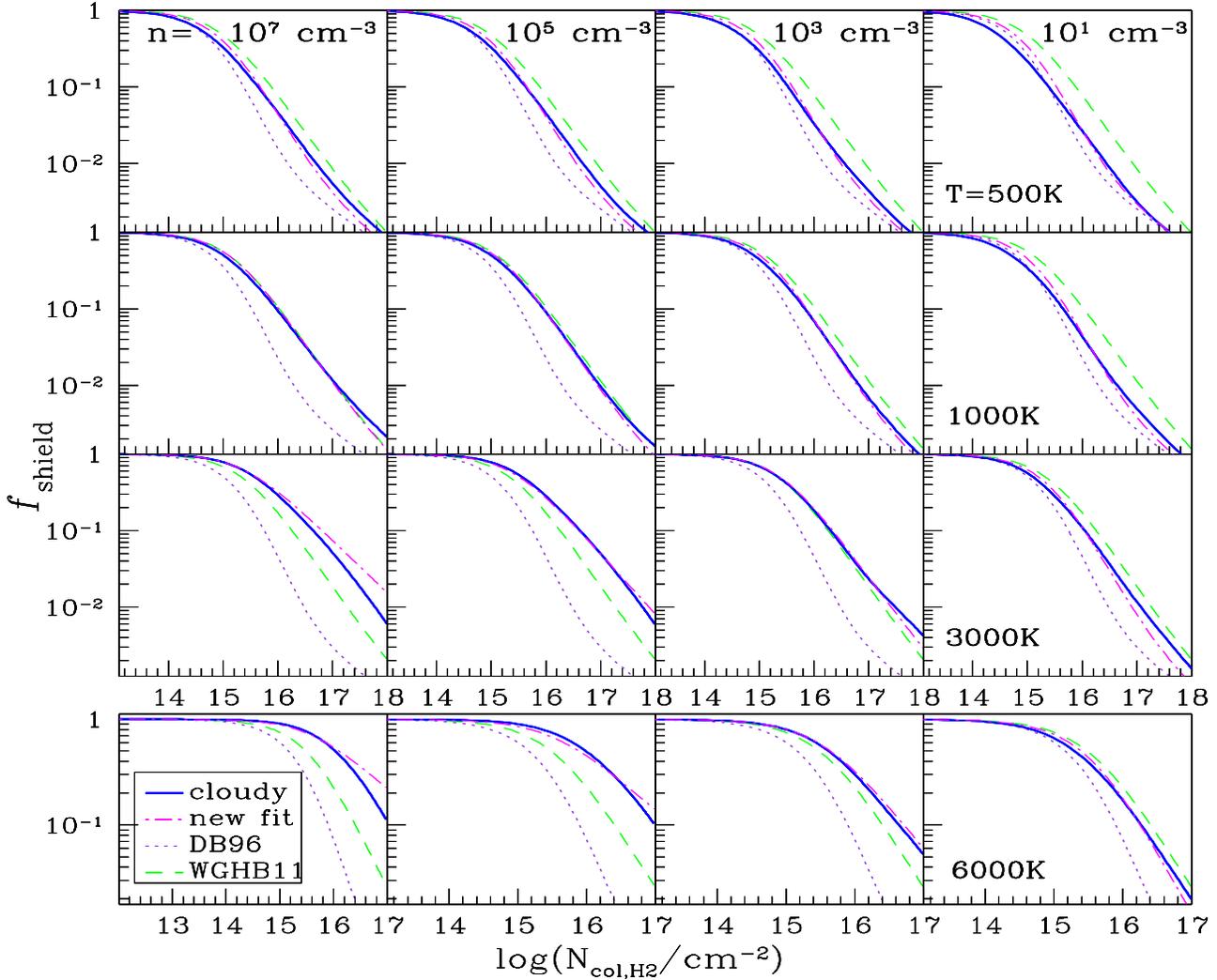}
  \caption{The optically--thick ${\rm H_2}$--photodissociation
rate is shown, parameterized by the self--shielding factor,
${\it f}_{\rm shield} (N_{H2},T) = k{\rm_{LW} (N_{H2},T)       
/ {\it k}{_{LW}(N_{H2}=0,T)}}$. Dark blue (solid) curves show 
results with fully--resolved rovibrational populations from 
\textsc{cloudy} models (at fixed density and temperature, 
as indicated on the figure). Green (dashed) and purple (dotted)
curves show the fitting formulae for $f_{\rm sh}$ provided by 
\citet{WGHB11} and \citet{DB96}, respectively. Magenta (dot-dash)
curves show results of the more accurate revised fitting formula 
in Equations \ref{Eq:NewFit}-\ref{Eq:NewAlpha}.}
  \label{Fig:FshFits}
\end{figure*}
\begin{figure*}
  \includegraphics[height=5.8in,width=7in]{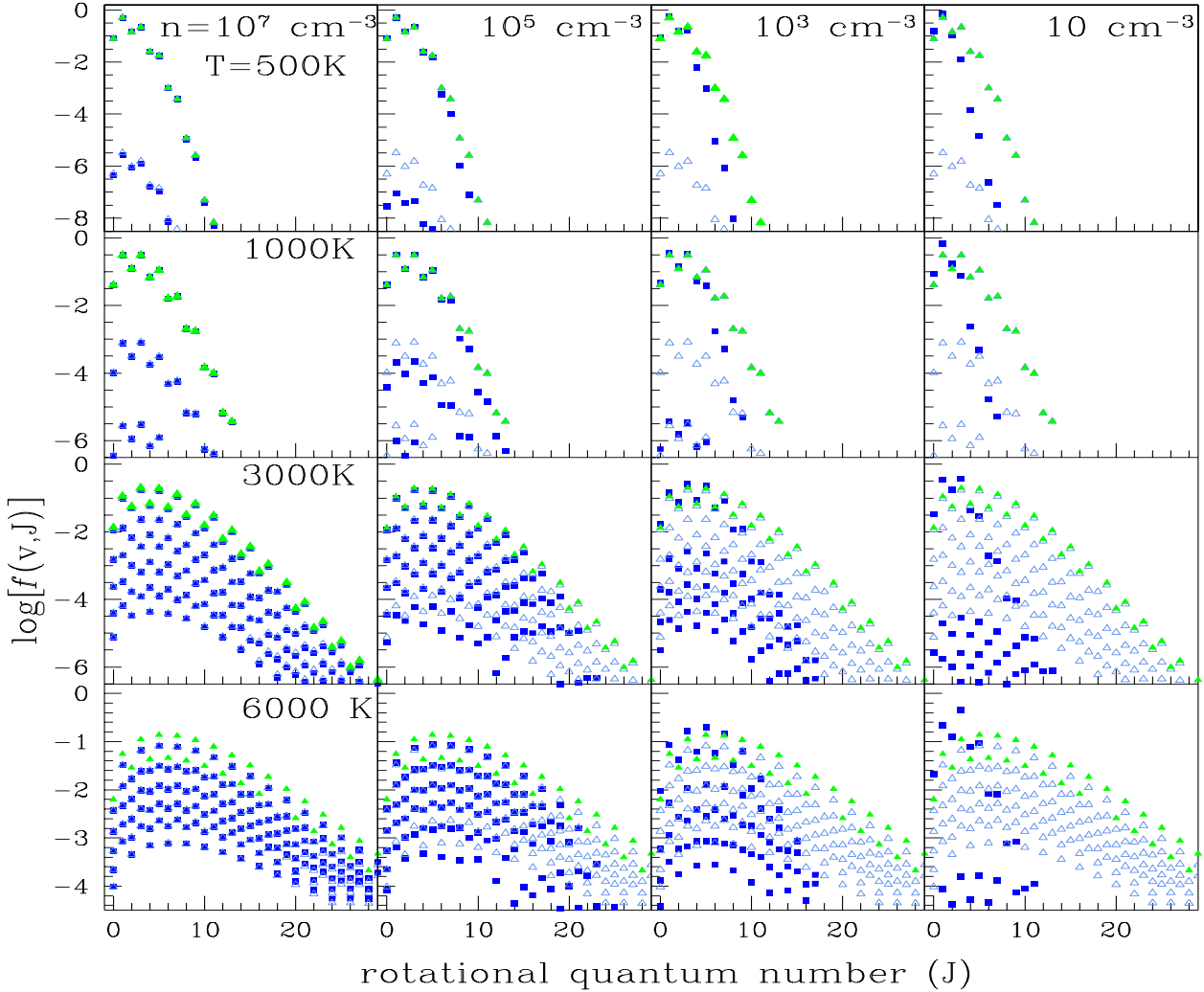}
  \caption{The fractional populations of ${\rm H_2}$ in our 
{\sc cloudy} models are shown for each rovibrational state (v<5,J). 
Dark blue squares are used for the fully--resolved (non-LTE)
results while grey (open) triangles are the full LTE populations. 
Green (filled) triangles show a Boltzmann distribution for
populations in the ground vibrational state only ${\rm v=0}$, 
which was the basis for a commonly used fitting formulas for
$f_{\rm shield}$ provided by \citet{WGHB11}.}
  \label{Fig:PopsPanels}
\end{figure*}

\begin{figure}
  \includegraphics[clip=true,trim=0.in 0.in 0.in 0.in,
    height=2.8in,width=3.2in]{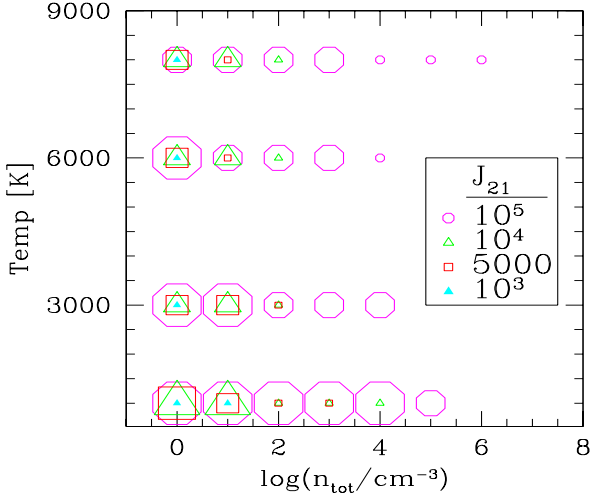}
  \caption{The densities and temperatures of {\sc cloudy} models 
in which the a strong UV flux increases the optically--thick 
${\rm H_2}$ photodissociation rate by more than a threshold factor 
$x$ are shown for three threshold values $x = 1.25,2,10$ with small, 
medium, and large (circles) respectively.  Results with four 
UV--intensities are shown, as indicated in the figure legend.}
  \label{Fig:BigDiffs}
\end{figure}
\begin{figure*}
  \includegraphics[clip=true,trim=0.0in 3.3in 0.in 0.in,
    height=3.3in,width=5.8in]{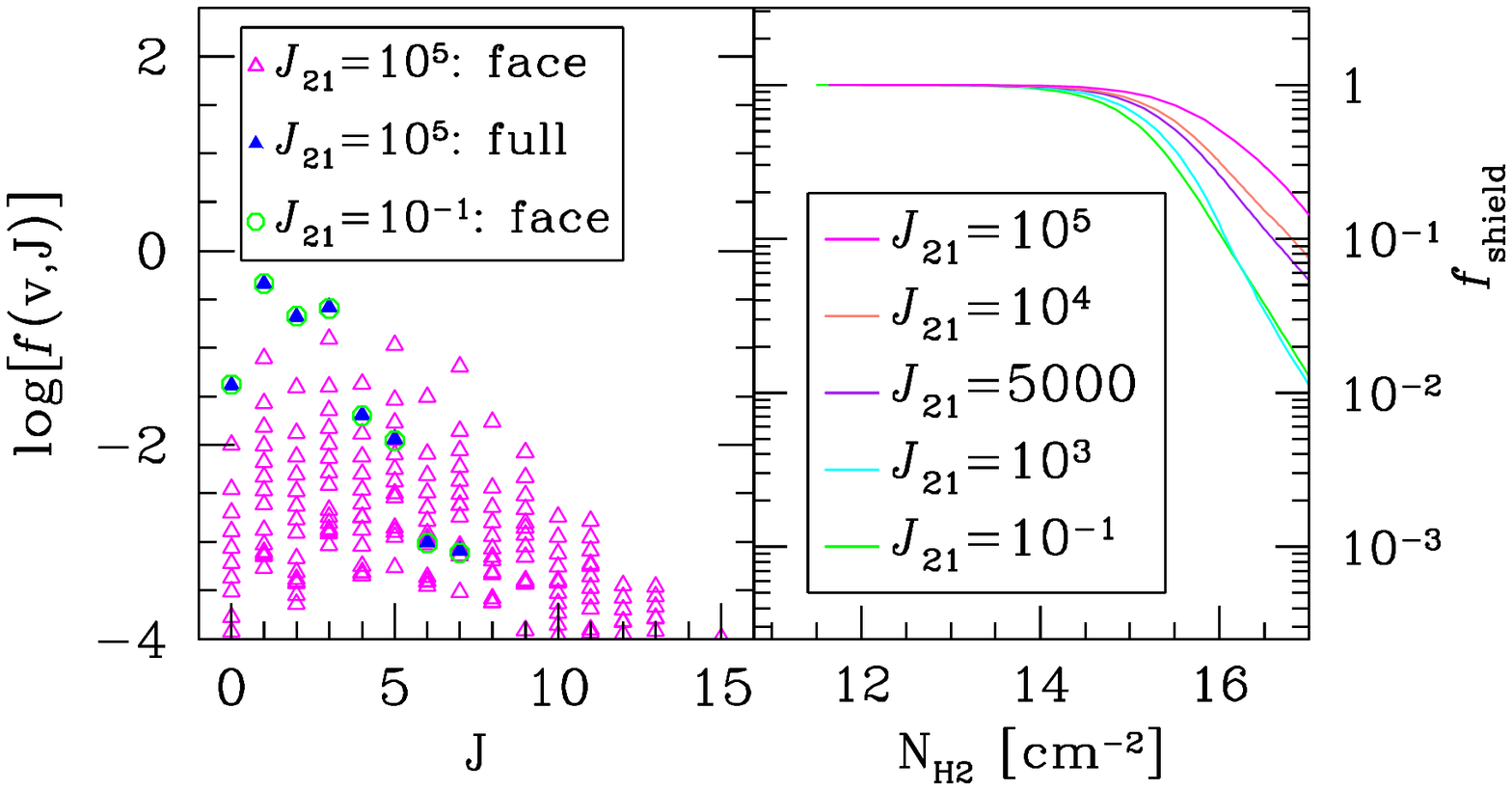} 
  \caption{{\bf Left}: The fractional population of ${\rm H_2}$ 
in rovibrational states (v,J) are shown for two different 
{\sc cloudy} models with the same temperature (${\rm T=6000}$K) 
and density (${\rm n=1~cm^{-3}}$). Blue (filled) triangles show 
the cumulative populations of a cloud with 
${\rm N_{H2}=10^{17}~cm^{-2}}$and irradiated by a strong--UV
flux ($J_{21}=10^5$, in the usual units). Magenta (open) 
triangles show only the populations at the illuminated face 
of the same cloud. Green (open) circles show the populations 
at the face of a cloud irradiated by a much weaker UV flux 
($J_{21}=0.1$). The panel shows that significant UV--pumping of 
excited rovibrational states occurs at the illuminated face, while 
the populations deeper in the cloud are unaffected by the strong 
incident UV flux. 
{\bf Right}: the optically--thick photodissociation rate, 
parameterized by a shield factor $f_{\rm shield}$, at varying 
UV flux levels (see legend). As expected for the strong--UV 
case ($J_{21}=10^5$), self--shielding is weaker due to pumping 
of ${\rm H_2}$ to a larger number of rovibrational states, while 
at $J_{21}=10^3$, the shielding is much closer to the 
``no--pumping''case ($J_{21}=0.1$). As in the left panel, 
the {\sc cloudy} models have constant temperature ${\rm T=6000}$K 
and density ${\rm n=1~cm^{-3}}$.}
  \label{Fig:PumpedPops}
\end{figure*}

\subsection{Cloudy Models}
\label{Subsec:CloudyModeling}

We use the most recent version of the publicly available 
package \textsc{cloudy} (v17.01) to calculate the 
rovibrational level populations of ${\rm H_2}$ in a gas 
of primordial composition and illuminated on one face 
(plane parallel geometry). We use a grid of  densities 
and temperatures in the range $n=10^{0-7}~{\rm cm^{-3}}$ 
and $T=500-10^5$K. We hold both constant in our models, 
in order to tease apart the effects on the rovibrational 
distribution and the resulting photodissociation rate. 

We use the \citet{DB96} galactic background spectrum for
the incident radiation in our fiducial models (see their 
Equation 23), but our results are not sensitive to this
choice. If we use the \citet{BvD87} interstellar radiation 
field instead, the photodisosciation is nearly unchanged. 
(However, neither of these spectra have prior processing
in the LW bands; \citealt{WGHB17} show that if the incident 
radiation originates from an older stellar population, 
absorption lines in the spectrum can change the resulting 
photodissociation rate by a factor of two or more.)  

We use the ``large ${\rm H_2}$'' model, details of which 
can be found in \citet{Shaw+05}; this model resolves all 
301 bound levels of the ${\rm H_2}$ electronic ground state 
and six excited electronic states. Several thousand energy 
levels of the molecule and approximately $5\times 10^5$ 
permitted transitions are included. In order to explore the 
results at high UV flux and low density, we decreased the 
threshold fractional abundance of ${\rm H_2}$ at which the 
resolved populations are calculated (``H2-to-H-limit'' in 
{\sc cloudy}) from its default ($10^{-8}$) to $10^{-12}$.

\subsection{The Rate Calculation}
Photodissociation of ${\rm H_2}$ occurs primarily via the 
two--step Solomon process, in which the molecule is first 
electronically pumped by a UV photon ($11.1-13.6$ eV) from 
the ground state ${\rm X}^1 \Sigma^+_g$ to the ${\rm B}^1 
\Sigma^+_u$ (Lyman) or ${\rm C}^1 \Pi_u$ (Werner) states. 
The subsequent decay to the vibrational continuum results 
in dissociation $\sim 15$ per cent of the time. The ``pumping 
rate'' from a given rovibrational state ($v,J$) to an excited 
electronic state ($v',J'$) is:
\begin{equation}
\zeta_ {\it{v,J,v',J'}} = \int_{\it{\nu_{\rm th}}}^{\nu_{13.6~eV}}4\pi
\sigma_{\nu}\frac{J_\nu}{h_{\rm P}\nu}{\rm d}\nu,
\end{equation}
where $\sigma_{\nu}$ is the frequency dependent cross--section
and $h_{\rm P}$ is Plank's constant. The frequency threshold,
$\nu_{\rm th}$, corresponds to the lowest energy photons capable 
of efficiently dissociating ${\rm H_2}$, with $h\nu= 11.18$ eV.
We do not include photons with energies above the Lyman limit,
which are likely to have been absorbed by the neutral IGM at 
the relevant redshifts (prior to reionization). In the direct 
collapse scenario with irradiation from a bright neighbor, the 
escape fraction of ionizing photons may be small anyway 
\citep[see e.g.][]{Wise+14}. Including ${\rm E > 13.6}$eV 
radiation would cause increased hydrogen ionization in the 
collapsing gas, which is not included in the present context. 

The dissociation rate from a given ($v,J$) is the product of 
the pumping rate and the fraction of decays that result in 
dissociation (summed over all excited states, $v',J'$): 
\begin{equation}
k_{{\rm diss},v,J} = \sum_{\it v',J'} \zeta_{\it v,J,v',J'}
{\it f}_ {{\rm diss},v',J'}.
\end{equation}
The dissociation probabilities $f_{{\rm diss},v',J'}$, are 
provided by \citealt{ARD00}. 

While {\sc cloudy} outputs the optically--thick dissociation 
rate itself, we re--calculate the rate in order to remove 
the effect of HI shielding ${\rm H_2}$ and isolate the 
self--shielding only. This also allows us to test the 
effects of changing one variable at a time, including 
the incident spectrum, the rovibrational distribution, 
and the temperature. WGHB11 found that the {\it total} 
shielding of ${\rm H_2}$ can be modeled with good accuracy 
by including a simple multiplcative factor that depends 
only on the HI column density:
$f_{\rm shield,tot}{\rm (N_{H2},N_{HI},T)} = 
f_{\rm shield}{\rm (N_{H2})} \times f_{\rm HI (N_{HI})}$.
Therefore, the effects we quantify for self--shielding 
are directly applicable to the total shielding, though HI
is not included in our fiducial calculations. 

We initiate the calculation with 
the \textsc{cloudy} incident spectrum at the irradiated 
face, and step through each discrete zone, summing the 
contributions to the frequency--dependent optical depth 
from the ${\rm H_2}$ transitions. In each zone, we use 
the resolved rovibrational populations from \textsc{cloudy}, 
${\it f}_{v,J}$. The total rate in a given zone is then:
\begin{equation}
k_{\rm diss} = \sum_{\it v,J} k_{{\rm diss},{\it v,J}} \times
{\it f}_ {v,J}
\end{equation}

\subsection{Critical Densities for LTE in ${\bf H_2}$ Rovibrational States}
\label{sec:ModelLevelPops}

One of the primary challenges in calculating the optically--thick 
photodissociation rate of ${\rm H_2}$ is determining the rovibrational 
level population distribution. In the non-LTE case, the distribution 
is time-dependent and sensitive to the gas temperature, density, as 
well as the UV pumping rate. 

In a low--density gas, collisional de-excitation is slow and the 
radiative decay rates entirely determine the cascade to lower states;
however, the radiative lifetimes are long enough that even at moderate
densities collisional de-excitation begins to have an effect
\citep{Drainebook}. The lowest energy states (${\rm v=0,J}$) reach 
LTE at critical densities of ${\rm n_{H} \lsim 10~cm^{-3}}$ at 
${\rm T=10^4K}$ (see Figure \ref{Fig:ncrit}). The critical 
density for a given transition is:
\begin{equation}
n_c \left(u \rightarrow l\right) = \frac{A \left(u \rightarrow l\right)}
{\gamma \left(u \rightarrow l\right)}
\end{equation}
here $\gamma \left(u \rightarrow l\right)$ is the collisional
de-excitation rate and ${A \left(u \rightarrow l\right)}$ is 
the spontaneous decay transition probability. Above the 
critical density, the fractional population of the state 
approaches its LTE value and is then dependent only on the 
temperature. Because ${\rm A \left(u \rightarrow l\right) 
\propto E_{lu}^5}$ (where ${\rm E_{lu}}$ is the energy of the
transition between upper and lower states), the critical 
density increases rapidly with J within a vibrational 
manifold while it decreases rapidly with temperature. Figure 
\ref{Fig:ncrit} shows the temperature dependence of the critical 
densities of the lowest energy levels (${\rm v=0-2}$,J). 
We use the fitting formulae for the collisional rates provided 
by \citet{LPF99}, which do not include all of the highest energy 
levels, but are nonetheless useful for the present purpose.

The left panel of Figure \ref{Fig:ncrit} shows the ${\rm H_2}$ 
fractional level populations (${\rm v=0-2}$,J) determined by 
{\sc cloudy} for a gas at T$=3000$K and ${\rm n=10^5~cm^{-3}}$.   
Resolved (non--LTE) populations (squares) diverge from the LTE
values (triangles) for states with energies above ${\rm E_{v,J}/k_B 
\sim 2\times 10^4}$K, for which the critical densities exceed 
${\rm n=10^5~cm^{-3}}$. In the right panel of this figure, we 
show the temperature dependence of the critical density for the 
same states ${\rm v=(0-2),J}$. From this figure, we see that 
the critical densities of the lowest $\sim 35$ states are less 
than ${\rm n=10^5~cm^{-3}}$ at ${\rm T=3000}$K, and the 
populations of these states (shown in the left panel) are 
indeed near their LTE values, as expected.

\subsection{Analytical Fitting Formulae for the Shielding Factor}
\label{Sec:AnalyticalFits}

Because of the computational expense of calculating the full
optically--thick ${\rm H_2}$--photodissociation rate, several
analytic fits have been suggested, parameterized by a 
``shielding factor:'' 
\begin{equation}
f_{\rm shield}{\rm (N_{H2},T)} = k_{\rm diss}{\rm (N_{H2},T)}/k_{\rm diss}
{\rm (N_{H2}=0,T)}.
\end{equation} 
\citet{DB96} (hereafter DB96) provided two useful fitting formulae 
for this purpose, the more accurate of which (their equation 37) 
has the form: 
\begin{multline}
f_{\rm shield,DB}\left(N_{\rm H_2}, T\right) =
\frac{0.965}{\left(1 + x/b_5\right)^\alpha}
+ \frac{0.035}{\left(1 +  x\right)^{0.5}}\\
\times \exp\left[-8.5 \times 10^{-4} \left( 1 +  x\right)^{0.5}\right].
\label{eq:DB37}
\end{multline}
DB96 set $\alpha=2, x \equiv N_{\rm H_2}/ 5 \times 10^{14}~
{\rm cm^{-2}}$, $b_5 \equiv b/10^5~{\rm cm~s^{-1}}$, where 
$b$ is the Doppler broadening parameter. 

WGBH11 showed that this fit is only accurate for low-density 
gas at temperatures of $\sim$a few hundred K. They modified it 
by setting $\alpha=1.1$ in order to better fit the population 
results in gas at $\sim$a few thousand Kelvin and densities 
$\sim 10^{3} {\rm cm^{-3}}$-- relevant for a gravitationally--
collapsing protogalactic halo. With this modification, 
self--shielding is weaker than the original DB96, appropriate 
for ${\rm H_2}$ populations that are spread out over more (v,J) 
states, as explained above. 

\section{Results and Discussion}
\label{Sec:Results}

\subsection{How accurate are fitting formulae for $\it {f_{shield}}$?}
\label{Sec:CloudyPops}

With our {\sc cloudy} models, we can evaluate the accuracy 
of the DB96 and WGHB11 fits compared to the ``true'' 
photodissociation rate over a wide swath of the parameter 
space. We include higher denities (${\rm n \leq 10^7 cm^{-3}}$)
and temperatures (${\rm T \leq 8000}$K) than in those previous 
studies, so that we can quantify the changes in the rate when 
the excited vibrational states become populated and eventually 
reach LTE. WGHB11 considered gas with density up to only 
${\rm n \lsim 10^3~cm^{-3}}$, and the DB96 fits were designed 
for much lower temperatures, ${\rm T \sim 100}$K. 

In Figure \ref{Fig:FshFits} we show our $f_{\rm shield}$ 
results in {\sc cloudy} models with low flux $J_{21}= 0.1$ 
(dark blue curves). The DB96 fitting formula (gray curves) 
significantly overestimates shielding compared to 
{\sc cloudy} at all but the lowest temperatures 
and densities (${\rm T=500K, n=10~cm^{-3}}$), as expected. 
The WGHB11 fit (green dashed curves) is more accurate up
to higher temperature and density ${\rm T\approx3000K}$, 
${\rm n \approx 10^{3}~cm^{-3}}$); however, it also gives
$f_{\rm shield}$ that is far too small (underestimates the
actual photodissociation rate) at ${\rm T>3000K}$, 
${\rm n > 10^{3}~cm^{-3}}$. While this regime may not be
relevant for the direct--collapse scenario, in which the 
gas thermodynamic evolution is determined at lower density 
${\rm n \sim 10^2}$, it could be important for simulations 
of ``Pop III'' star formation in the presence of a strong 
incident UV flux.   

Figure \ref{Fig:PopsPanels} shows the origin of the 
discrepancy of the WGHB11 fit. The resolved level populations 
for each of the \textsc{cloudy} models (dark blue squares) 
are shown in comparison to the predicted LTE populations 
(light blue triangles).  The WGHB11 fit was based on a 
Boltzmann distribution {\it only in the ground vibrational 
state}, shown in the figure by green triangles. This $v=0$ 
model is a good approximation up to ${\rm T\sim 3000}$K.
At higher temperatures and densities, however, the populations
in $v>0$ states increase. Because the number of populated states 
is what matters for $f_{\rm shield}$ -- a greater number of 
populated states leads to lower effective column density for 
shielding -- the $v=0$ assumption for $f_{\rm shield}$ is 
erroneously small. 

\subsection{A New Fitting Formula for $\it {f_{\rm shield}}$} 

In order to find a more accurate fitting formula for 
$f_{\rm shield}$, we need to increase the sensitivity 
to temperature for densities above ${\rm n > 10~cm^{-3}}$, 
reflecting the strong temperature dependence of the 
critical densities. 

We make use of the original form provided by DB96, which
WGHB11 modified with a new factor $\alpha$. Here, we introduce
a density and temperature dependence of $\alpha$, so that the
effect of spreading populations over more levels at high n/T
is captured: 
\begin{multline}
f_{\rm shield}\left(N_{\rm H_2}, T\right) =
\frac{0.965}{\left(1 + x/b_5\right)^{\alpha(n,T)}}
+ \frac{0.035}{\left(1 +  x\right)^{0.5}}\\
\times \exp\left[-8.5 \times 10^{-4} \left( 1 +  x\right)^{0.5}\right].
\label{Eq:NewFit}
\end{multline}
\begin{multline}
\alpha {\rm \left(n,T\right) = A_1(T) \times 
 dexp\left(-c_1 \times log\left(n/cm^{-3}\right)\right) + A_2(T)}\\
{\rm A_1(T) = c_2 \times log\left(T/K\right) - c_3}\\
{\rm A_2(T) = -c_4 \times log\left(T/K\right) + c_5}\\
\label{Eq:NewAlpha} 
\end{multline}

The best fit parameters, optimized using the {\sc amoeba} routine 
in Numerical Recipes \citep{NRecipes}, are as follows: 
$c_1=0.2856,~c_2=0.8711,~c_3=1.928,~c_4=0.9639,~c_5=3.892$. 
The accuracy of the new fit is improved over the entire parameter
space compared to the previous fits, as shown in Figure 
\ref{Fig:FshFits}, with typical errors of a few percent for
$f_{\rm shield}$ up to ${\rm N_{H2} = 10^{17}~cm^{-2}}$. 

It is worth noting that \citet{Richings+14} also used {\sc cloudy} 
models similar to ours to investigate the accuracy of various 
fitting formulae compared to the ``true'' optically--thick rate. 
However, in that study, the true rate is calculated by {\sc cloudy}
itself, which includes HI--shielding of ${\rm H_2}$ along with 
self--shielding. Therefore, their results are particular to those 
{\sc cloudy} models and their specific HI/${\rm H_2}$ profiles.
As a result, we have not compared our results to the fitting
formula they provide.

\subsection{Is pumping important at the relevant flux strength?}
\label{Sec:Pumping}

In order to quantify the effect of a strong incident UV flux 
on the optically--thick ${\rm H_2}$--photodissociation rate 
in a primordial cloud, we have run a grid of {\sc cloudy} 
models with a range of flux intensities, $J_{21}=\left(0.1, 
10^5\right)$, gas temperatures, ${\rm T=(500,8000)K}$, and 
densities ${\rm n = 10^{0-7}~cm^{-3}}$. In Figure 
\ref{Fig:BigDiffs} we show the combinations for which a 
strong flux changes the shielding by more than a threshold factor 
$f_{\rm shield}\left(J_{21}\right)/f_{\rm shield}\left (J_{21} 
= 0.1 \right) > x$, for three threshold values $x = 1.25,2,10$. 
Our results are shown if the above criterion is fulfilled at 
any column density up to ${\rm N_{H2} = 10^{17}~cm^{-2}}$. 

In the majority of cases, there is no signficant change in 
the self-shielding except at very high intensity, 
$J_{21}=10^5$. However, at low densities ${\rm n \leq 10^2
~cm^{-3}}$, even a ``moderate'' UV intensity increases the
dissociation rate, For example, an incident flux of only
$J_{21}=5,000$ reduces shielding by a factor $>2$ for a gas  
${\rm n = 10~cm^{-3}}$ and by more than an order of magnitude 
when ${\rm n= 1~cm^{-3}}$. Even a flux of just $J_{21}=10^3$ 
leads to a $40-50$ per cent decrease in the shielding in the 
lowest density cases ${\rm n \leq 10~cm^{-3}}$. 

Why is there a pronounced difference in self--shielding in 
the low--density, high--flux cases? Figure \ref{Fig:PumpedPops} 
shows the origin of this effect. The fractional populations 
of the ${\rm H_2}$ rovibrational levels are shown, in the 
left--hand panel in both the strong-- and weak--UV cases, 
for a low density {\sc cloudy} model (${\rm n = 10~cm^{-3}}$).
Because pumping has the greatest effect near the edge of
the cloud, where there is the least shielding, we show the
level populations {\it at the irradiated face}. In the 
strong--UV case (magenta open triangles), many excited states 
${\rm v >0}$ populated. This is to be expected for a low--density 
gas, wherein strong UV pumping leads to a radiative cascade 
through excited states that are not populated by collisional 
processes alone. In contrast, in the weak--UV case, (green circles) 
only a few states are populated. We also show the cumulative 
populations (blue filled triangles) in the strong--UV case, which 
are nearly identical to those with weak--UV. This illustrates 
that pumping does not affect the bulk populations, as is expected 
due to shielding.

As discussed above, the spreading out of ${\rm H_2}$ column 
density in more (v,J) states by pumping decreases the effective 
self--shielding column density, since each molecule ``sees'' fewer 
molecules in the same (v,J) state between it and the irradiated 
face of the cloud. The right panel of Figure \ref{Fig:PumpedPops} 
shows the ``pumped'' shielding factor $f_{\rm shield}$ 
($J_{21} = 5,000$) is three times larger than for small $J_{21}$ 
at $N{\rm_{H2} = 10^{17}~cm^{-2}}$. With $J_{21}=10^5$, the 
optically--thick photodissociation rate is $\sim$ an order of 
magnitude larger than the same for small $J_{21}$.
At $J_{21} = 10^3$, the largest change in the shield factor is 
$\sim 40$ per cent at ${\rm N_{H2}} \sim$ a few $\times 10^{15} 
{\rm cm^{-3}}$. In both panels of Figure \ref{Fig:PumpedPops}, 
all of the {\sc cloudy} models have constant ${\rm T=6000}$K and 
${\rm n = 1~cm^{-3}}$.

In higher density models, ${\rm n>10^3~cm^{-3}}$, there 
is no pumping effect on the ${\rm H_2}$ rovibrational 
populations. This is because the populations in the 
first few vibrational states (the most important for 
self--shielding) are already tending toward (or in) LTE at
these densities (see Figure \ref{Fig:PopsPanels}). In addition, 
larger neutral hydrogen column densities likely decrease the 
UV pumping more in these higher density cases, even near the 
illuminated face of the cloud. 

It is worth noting that the density/temperature/$J_{21}$ 
space where pumping modifies the self--shielding behavior 
is very similar to that relevant for determining the critical 
flux to keep the halo ${\rm H_2}$--poor, and thus for 
potential direct collapse to a supermassive--black hole seed. 
For example, \citet{O01,SBH10}, show that the bifurcation 
in the cooling history of DCBH candidate halos occurs when 
collapsing gas reaches ${\rm n \sim 10^2~cm^{-3}}$ and 
${\rm T=8000}$K and the critical flux is $J_{21}\sim 10^{3-4}$ 
(WGBH11 and references therein). Therefore, it is likely 
that weaker shielding caused by UV--pumping at densities 
${\rm n = 10^{0-2}~cm^{-3}}$ is indeed relevant in the 
direct--collapse case, and pumping may well lead to a 
{\it smaller} $J_{\rm crit}$ if accounted for in simulations. 

\section{Conclusions}
\label{Sec:Conclusions}

Using {\sc cloudy} non-LTE models of pristine gas, we 
calculate the optically--thick ${\rm H_2}$ photodissociation 
rate using the resolved level populations and compare to the 
fitting formulae most commonly used in simulations. We find 
that the formula provided by \citet{WGHB11} is most accurate
at moderate densities, $n\lsim 10^{3}~{\rm cm^{-3}}$, but 
fails to capture the weakening of shielding at higher densities 
and temperatures, when the populations in ${\rm v>0}$ tend to 
LTE. We provide a new modification to the fitting formula that 
increases its accuracy at all densities and temperatures and is 
a good fit up to ${\rm n \sim 10^{7}~{\rm cm^{-3}}}$, T$=8000$K, 
${\rm N_{H2} = 10^{17}~cm^{-2}}$. This new analytical fit can 
be easily used in simulations and one--zone models to better 
approximate the optically--thick photodissociation rate.  

We also find that the photodissociation rate can be 
significantly increased in the presence of a strong UV flux, 
$J_{21} \gsim$ a few $\times 10^3$ due to pumping of molecules 
to excited rovibrational states. Increasing the number of 
populated states decreases the effective self--shielding column 
density that each molecule ``sees,'' and thus increases the 
optically--thick rate. This effect occurs only in gas at 
relatively low densities,  ${\rm n \leq 10^{2}~{\rm cm^{-3}}}$, 
which happen to be similar to those important for the 
determination of $J_{\rm crit}$ for direct--collapse black hole 
formation. We find shielding is decreased by as much as an order 
of magnitude in some cases for an incident flux 
$J_{21} \sim 10^4$, and even a flux as low as $J_{21} \sim 10^4$ 
can cause a change $>40$ per cent in some cases. 

\section*{Acknowledgments}

We thank Gargi Shaw for assistance with the ``Large ${\rm H_2}$''
model in {\sc cloudy} and Greg Bryan for useful discussions. 
ZH acknowledges support from NASA grant NNX15AB19G. JWG 
acknowledges support from the NSF Graduate Research Fellowship 
Program.

\bibliography{paper}

\label{lastpage}

\end{document}